\documentclass[article,prl,twocolumn,showpacs]{revtex4}
	\usepackage{graphicx,amsmath,amssymb,color}
	\usepackage[utf8]{inputenc}
	\usepackage{mathrsfs}
	\usepackage{amsfonts}
	\usepackage{caption}
	\usepackage{enumerate}
	\usepackage[toc,page]{appendix}
	\usepackage{url}
	\usepackage[normalem]{ulem}
	\usepackage[ocgcolorlinks,colorlinks=true,linkcolor=blue,citecolor=red]{hyperref}
	\usepackage{latexsym}
	\usepackage{amsthm}
	\usepackage{verbatim}
	\usepackage{psfrag}
	\usepackage{fancyhdr}
	\usepackage{pdflscape}
	\usepackage{lipsum}

\begin{document}
		\title{The Magical Number Seven: An Unexpected Dimensional Threshold in Quantum Communication Complexity}
		\author{Armin Tavakoli$^{1,2}$, Marcin Paw\l{}owski$^2$, Marek \.Zukowski$^2$, Mohamed Bourennane$^1$}
		\affiliation{$^1$Department of Physics, Stockholm University, S-10691 Stockholm, Sweden.\\ $^2$Institute of Theoretical Physics and Astrophysics, Uniwersytet Gda\'nski, PL-80-952 Gda\'nsk, Poland.}

		
		\date{\today}
		
		
		\begin{abstract}
			Entanglement-assisted classical communication and transmission of a quantum system are the two quantum resources for information processing. Many information tasks can be performed using either quantum resource. However, this equivalence is not always present since entanglement assisted classical communication is known to sometimes be the better performing resource. Here, we show not only the opposite phenomenon; that there exists tasks for which transmission of a quantum system is a more powerful resource than entanglement assisted classical communication, but also that such phenomena can have a surprisingly strong dependence on the dimension of Hilbert space. We introduce a family of communication complexity problems parametrized by dimension of Hilbert space and study the performance of each quantum resource. We find that for low dimensions, the two resources perform equally well, whereas for dimension seven and above, the equivalence is suddenly broken and transmission of a quantum system becomes more powerful than entanglement assisted classical communication. Moreover, we find that transmission of a quantum system may even outperform classical communication assisted by the stronger-than-quantum correlations obtained from the principle of Macroscopic Locality.

		\end{abstract}
		
		
		
		\maketitle
		
		{\it Introduction.---} Information processing can be enhanced beyond classical limitations by using quantum resources. For instance, parties can share an entangled systems on which measurements are performed and the non-classically correlated outcomes used to enhance classical communication for sake of e.g. transferring or hiding of  information.  Alternatively, the quantum resource can be the preparation and transmission of a quantum system, from which information is extracted by performing  suitable measurements. These two quantum resources: entanglement-assisted classical communication (entanglement assisted classical communication) and transmission of a quantum system (transmission of a quantum system), have a deep but to great extent unexplored relationship. Many communication tasks can be performed with either resource, and they typically share several qualitative and quantitaive properties.

		For instance, the first quantum key distribution protocol,  BB84  \cite{BB84}, is based on transmission of a quantum system. Still, it can  be translated to  an entanglement assisted classical communication protocol \cite{BBM92}, in which the measurements on the entangled state directly correspond to the measurements and state preparations in  BB84. Due to this similarity, the security of the former derives from relating it to the latter \cite{SP00}. There are several other examples  \cite{YLH08, TI15, FGM01, TC15} of such qualitative equivalence between the two quantum resources.

		A more quantitative relation between  entanglement assisted classical communication and transmission of a quantum system manifests itself in communication complexity problems. In such problems two, or more partners, hold data known only to each of them, but one of them has to give the value of a function which depends on the full data set of all of them. If there is a communication constraint, which does not alllow transmission of all data needed to give the value of the function,
		this constitutes an example of a nontrivial communication complexity problem. 
		
		In communication complexity problems, one is interested in either finding a minimal amount of communication required to compute a function with inputs distributed between a group of parties, or to find the maximal probability of computing the function given limited communication. For instance, Ref. \cite{BCD01} considers an example of the latter. Two parties Alice and Bob, each hold random bits $(x_0,x)$ and $y$ respectively. Their aim is to let Bob compute the value of the function $f\!=\!(-1)^{x_0+xy}$ while Alice communicates only one (qu)bit. The communication complexity problem is linked with the CHSH inequality \cite{CHSH69},  as the inequality can be put in such a way that gives the upper bound on classical strategies for the average of the product of the answer $g$ with the actual value of the function $f$, that is $\langle gf\rangle$. For all classical protocols one has  $\langle gf\rangle\!\leq\! \frac{1}{2}$, see \cite{BZPZ04}, that is the probablitiy to get a correct answer is $3/4$. If the parties share a maximally entangled two-qubit state, they can perform measurements,  the local  settings of which depend on $x$ and $y$, respectively, such that the CHSH inequality is maximally  violated. If Alice communicates to Bob  one bit of information, dependent on her $x_0$ and the value obtained in the measurement, Bob can compute $f$ with a probability beyond the classical bound, that is  $1/2\!+\!1/2\sqrt{2}$. However, the same success probability can be achieved if Alice sends a qubit \cite{G01, TROJEK}. The protocol is such that the state of the qubit depends on $x_0$ and $x$. Bob makes a measurement whose setting depends on $y$.  In this protocol the better-than-classical can be linked to violation of constraints of classical dimensionality \cite{M-expDW}.

		From both a conceptual and applied point of view, it is important to ask whether entanglement assisted classical communication and transmission of a quantum system are always equivalent resources in communication tasks. In Ref. \cite{PW12}, it was shown that in binary-outcome communication complexity problems with communication restricted to one (qu)bit, entanglement assisted classical communication is always at least as good as transmitting a qubit. Examples are known for which entanglement assisted classical communication is strictly better than a broad class of known transmission of a quantum system strategies \cite{PZ10} and it can be shown using the method of Ref. \cite{NV15} that entanglement assisted classical communication is strictly better than any possible transmission of a quantum system strategy for particular tasks considered in Ref. \cite{PZ10}.
		
		We will show the opposite phenomenon. We introduce a family of specific communication complexity problems and study them with entanglement assisted classical communication and transmission of a quantum system respectively. From Refs. \cite{BZPZ04, TZ16} we know that every Bell inequality can be turned into a communication complexity problem. The communication complexity problems of our study  will be linked with the so called CGLMP Bell-type inequalities \cite{CGLMP02}. The reason we consider these games linked to the CGLMP inequalities is that these inequalities define the only known faces of the local polytope describing all possible classical correlations for two observers who have two settings and more than two possible outcomes. This is arguable the most interesting scenario to analyze the comparative strength of classical communication assisted by entangled states by which the CGLMP inequalities admit large violations. The games we introduce are generalizations of the  communication complexity problem corresponding to $d=3$ considered in Ref. \cite{BZZ02}. Again one can show that a Bell experiment in which the CGLMP inequality is maximally violated, can be used to breach the classical constraints of the communication complexity problem. An alternative solution of the communication complexity problems will be also formulated, such that  Alice transmits a $d$ dimensional quantum system to Bob. We will analyze and compare the performance of the two quantum resources in case of the family of communication complexity problems. Surprisingly, both quantum protocols are equally effective for $d\leq 6$ but for $d\geq7$, the second quantum protocol outperforms its entanglement-assisted analog. 
		
		Communication complexity problems can be viewed as games in which the parties are awarded a number of points depending on which outcomes they return conditioned on the inputs they are given.
	In general, given that all inputs are equiprobable, a two-party Bell inequality can be written as
		\begin{equation}
		\sum_{a,b,x,y}c_{a,b,x,y}P(a,b|x,y)\leq C,
		\end{equation}
		where $x$ and $y$ stand for the inputs while $a$ and $b$ for the outputs of the parties. We can regard the coefficients $c_{a,b,x,y}$ as payoffs in a nonlocal game \cite{W10}, awarded to the parties when, upon receiving inputs $x,y$ they output $a,b$.  Such nonlocal games can be turned into games in which one party, say Alice, sends a message, based on her local data, to Bob who recieves a payoff depending on whether he manages to call out the value of some function depending on $a,b,x,y$. A quantum version of this game relies on that entanglement can in principle violate the classical bound $C$. The amount of communication in the game is constrained to the logarithm of the size (denoted by $d$) of the alphabet of the outcomes of Bob. To cast the game in terms of transmission of a quantum system, we keep both the payoffs and the capacity of the channel as above, but instead of sharing entanglement and communicating classically the parties transmit a quantum system of dimension $d$. There are many known examples of such mappings of nonlocal games into communication based games leading to the same average payoffs \cite{PZ10,M-expDW,LPM13,MLP14}. In this work we apply generalizations of such ideas to a game related to the CGLMP inequality.

		{\it The CGLMP inequality.---} 
		The CGLMP inequality constitutes a bound for classical correlations in bipartite Bell scenarios in which Alice and Bob respectively choose one of two measurements, $A_x$ and $B_y$ for $x,y\in\{0,1\}$, returning one of $d$ possible outcomes $a,b\in\{0,...,d-1\}$ respectively. As the probability for each pair of inputs is $1/4$, the inequality can be written,
		\begin{multline}\label{CGLMPexpr}
		I_d\equiv \frac{1}{4}\sum_{k=0}^{\lfloor\frac{d}{2}\rfloor-1} c_k
		\Big[\Big(P_{A_0B_1}(b=a+k)+P_{A_0B_0}(a=b+k)\\ +P_{A_1B_1}(a=b+k)+P_{A_1B_0}(b=a+k+1)\Big)\\ -\Big(P_{A_0B_1}(b=a-k-1)+P_{A_0B_0}(a=b-k-1)\\+P_{A_1B_1}(a=b-k-1)+P_{A_1B_0}(b=a-k)\Big)\Big]\leq \frac{1}{2},
		\end{multline}
		where $c_k=1-\frac{2k}{d-1}$ and $P_{A_xB_y}$ denotes probability conditioned on Alice's and Bob's measurements being $A_x$ and $B_y$ respectively.

		{\it A game related to the CGLMP inequality.---}
		 We will now tailor a communication complexity problem to the CGLMP inequality for any $d$. To this end, let a referee provide Alice and Bob each with random and uniformly distributed input data; Alice with $\{x_0,x\}$ and Bob with $y$, such that $x,y\in\{0,1\}$ and $x_0\in\{0,...,d-1\}$ for some known $d\geq 2$.  The communication complexity problem is as follows. Alice is allowed to communicate to Bob, at most $\log d$ bits of information which depending on the class of the communication complexity problem is either classical information or a quantum $d$-dimensional system. Evidently, Alice cannot communicate all data $(x_0,x)$ to Bob. Basing on the information received from Alice, local data and local actions, Bob must output a number $G\in\{0,...,d-1\}$. The communication complexity problem is such that the partnership earns $c_k$ points if $G$ is the value of the function
		\begin{equation}\label{fk}
		f_k(x_0,x,y)=x_0-xy-(-1)^{x+y}k \mod{d}
		\end{equation}
		where $k=0,...,\lfloor \frac{d}{2}\rfloor-1$, and is penalized by losing $c_k$ points if $G$ is the value of the function
		\begin{equation}\label{hk}
		h_k(x_0,x,y)=x_0-xy+(-1)^{x+y}(k+1) \mod{d}.
		\end{equation}
		The functions $f_k$ and $h_k$ are chosen such that the referee can always pinpoint whether Bob has returned the value of either $f_k$ or $h_k$ or none of them. Note that at most one $k$ applies to given answer of Bob. For simplicity, we will label the introduced communication complexity problem as $\mathbb{G}_d$.
		
		For example, if we set $d\!=\!2$ we have only $k\!=\!0$, and $f_0\!=\!x_0-xy \mod{2}$ while $h_0\!=\!f_0+1\mod{2}$. Here, we recognize $f_0$  as the objective function $f$ in the communication complexity problem based on the CHSH inequality discussed earlier. Thus whenever the partnership succeed in the communication complexity problem they earn $c_0\!=\!1$ points, while losing $c_0\!=\!1$ points whenever they  fail which corresponds to computing $h_0$.

		Generally, the average number of points earned by the partnership is
		\begin{equation}\label{taskfunction}
		\Delta_d\equiv \sum_{k=0}^{\lfloor \frac{d}{2}\rfloor-1}c_k\left[P(G=f_k)-P(G=h_k)\right].
		\end{equation}
		
		The value of $\Delta_d$ depends on the explicit strategy undertaken by the partnership. We shall
			limit
			considered
			strategies to such in which the Bob's guess is
			$G=m(x_0,x)-b(y)\mod{d}$, where $m(x_0,x)$ is the message of Alice, which depends on her data. We shall call such strategies linear.
			As
			it is shown in Ref. \cite{TZ16}
			in such 
			a case the optimal message of Alice has the form of $m(x_0,x)=x_0+a(x)\mod{d}$,
			and the value of $\Delta_d$ is 
			equal to the bound of the associated Bell inequality.

		With such a linear strategy, we can show that $\mathbb{G}_d$ can be related to the CGLMP inequality. We first compute $P(G=f_k)$ which is equivalent to $P\left(-(-1)^{x+y}k-xy=a-b\right)$. Since the distributions of $x,y$ are uniform, we expand $P(G=f_k)$ as
		\begin{multline}
		P\left(G\!\!=\!\!f_k\right)=\frac{1}{4}\left(P_{A_0B_1}(b=a+k)+P_{A_0B_0}(a=b+k)\right. \\ \left. +P_{A_1B_1}(a=b+k)+P_{A_1B_0}(b=a+k+1)\right).
		\end{multline}
		Similarly, we expand $P(G=h_k)$ or $P\big((-1)^{x+y}(k+1)-xy=a-b\big)$ as
		\begin{multline}
		P\left(G\!\!=\!\!h_k\right)=\frac{1}{4}
		\Big(P_{A_0B_1}(b=a-k-1)+P_{A_0B_0}(a=b-k-1)\\
		+P_{A_1B_1}(a=b-k-1)+P_{A_1B_0}(b=a-k)\Big).
		\end{multline}
		Using the above two expressions for $P(G=f_k)$ and $P(G=h_k)$, the quantity $\Delta_d$ in \eqref{taskfunction} can be written in exactly the same form as the CGLMP inequality \eqref{CGLMPexpr}. Thus,  $\max \Delta_d=\max I_d$.

		{\it $\mathbb{G}_d$ with entanglement assisted classical communication. ---} Let Alice and Bob perform measurements on a shared entangled state and then Alice communicates $\log d$ bits of information. Alice (Bob) uses her (his) input $x$ ($y$) to determine one of two possible local measurements $A_{x}$ ($B_{\overline{y}}$, where the bar denotes bit-flip operation and is only introduced for simplifying purposes), performed on the entangled state. The measurements of Alice and Bob return outcomes denoted $a_{x},b_{y}\!\in\!\{0,...,d-1\}$ respectively. Finally, Alice sends a message $m=a_x+x_0\mod{d}$ to Bob. Bob's bet for the value is $G=m-b_{y}\mod{d}$. The optimality of such a message in the entanglement assisted classical communication protocol under the assumption of $G=m+b\mod{d}$ is shown in Ref. \cite{TZ16}.
		 Since quantum theory can violate the CGLMP inequality, we can go beyond the limitation  $\Delta_d\leq 1/2$. In table \ref{tab:1}, we have given the optimal values of $\Delta_d$ for $d=2,...,11$. For $d=2,...,8$ the maximal violations of the CGLMP can be found in \cite{ADGL02}. They can be verified by use of the second level of the Navascues-Pironio-Acin (NPA) hierarchy of quantum correlations \cite{NPA07}. For $d=9,10,11$, we have bounded $I_d$ from above using the intermediate level, $Q_{1+ab}$, of the NPA hierarchy and we have proved optimality by reproducing the upper bounds by interior point optimization using semidefinite programs (SDPs) \cite{VB96}. The performance of entanglement-assisted strategies in $\mathbb{G}_d$ improves with increasing $d$.
		
		
		{\it $\mathbb{G}_d$ with transmission of a quantum system. ---} Let us study the performance of the second quantum resource, transmission of a quantum system,  in $\mathbb{G}_d$.  
		In such a scheme, in each round of the game Alice is constrained to sending a $d$ dimensional state to Bob as her message. Alice's $2d$ possible inputs $(x_0,x)$ must be one-to-one linked with her state preparation $|\psi_{x_0x}\rangle$. Bob will use his binary input $y$ to define the measurement basis in which he measures the arriving system in one of the states $|\psi_{x_0x}\rangle$. He obtains an outcome $b_{y}\in\{0,...,d-1\}$, with probability which we denote as  $P(b_{y}
		|y, \psi_{x_0,x})$. Bob puts $G=b_{y}$.	Specifically, Bob will output $G=f_k$ (or in the case of bad luck, $G=h_k$) if  $b_{y}=x_0-(-1)^{x+y}k-xy\mod{d}$ (respectively, he would get $h_k$ if $b_{y}= x_0-xy+(-1)^{x+y}(k+1)\mod{d}$). Therefore, the partnership's average earned points $\Delta_d$, which is given by (\ref{taskfunction}), acquires the following specific form:
		\begin{eqnarray}\label{racc}
			&\Delta_d 
			= \frac{1}{4d}\sum_{k=0}^{\lfloor \frac{d}{2}\rfloor -1}c_k \sum_{x_0,x,y}& \nonumber \\
		&\times	\Big(P(b_{y}\!\!=\!\! f_k\!\!-\!\!x_0 |y,\psi_{x_0x})-P(b_{y}\!\!=\!\!h_k\!\!-\!\!x_0|y, \psi_{x_0x} )\Big) \nonumber&\\
		\end{eqnarray}
		Note, that Alice and Bob can use in their protocol whichever (pure) set of  $d$-dimensional states,  related to Bob's data. However, only a specific choice of such a set can give the maximal value of $\Delta_d$.

			\begin{table}[t]
				\centering
				\begin{tabular}{|c|c|c|c|c|}
					\hline
					$d$ & $ \text{lower bound } \Delta^{QS}_d $ & $\Delta^{Ent}_d$ &  $\Delta^{ML}$ & $\Delta^{QS}_d-\Delta_d^{Ent}$ \\ [0.5ex]
					\hline
					2 &  0.7071 & 0.7071 & 0.7071 &  0 \\
					3 &  0.7287 & 0.7287 & 0.7887 & 0\\
					4 & 0.7432 & 0.7432& 0.8032 & 0\\
					5 &  0.7539 & 0.7539 & 0.8249 & 0\\
					6 &  0.7624 & 0.7624 & 0.8345 & 0\\
					7 & 0.7815 & 0.7694 & 0.8461 & 0.0121\\
					8 &  0.8006 & 0.7753 & 0.8529 & 0.0253\\
					9 &  0.8622 & 0.7804 & 0.8605 & 0.0818\\
					10 & 0.8778 & 0.7849 & 0.8657 & 0.0929\\
					11 & 0.8870 & 0.7889 & 0.8713 & 0.0981\\
					\hline
				\end{tabular}
				\caption{The performance of entanglement assisted classical communication, transmission of a quantum system and Macroscopic Locality in $\mathbb{G}_d$ for some low values of $d$.}
				\label{tab:1}
			\end{table}

		 {\it Maximal value of $\Delta_d $ for transmission of a quantum system.--- } To search for the optimal strategy using transmission of a quantum system, that is for the optimal set of states $|\phi_{x_0,x}\rangle$, and measurement bases dependent on $y$, which maximize  $\Delta_d $ of (\ref{racc}), we have used SDPs in a see-saw method \cite{WW01, PV10}, iterating between running an SDP over the state preparations of Alice, and then over the measurements of Bob. Such a procedure performs optimization from within the set of valid quantum states and measurements, and is therefore not guaranteed to return a globally optimal value of  $\Delta_d$. The see-saw method only provides a lower bound on $\Delta_d$. In table \ref{tab:1} we compare the performance in $\mathbb{G}_d$ by means of transmission of a quantum system ($\Delta_d^{QS}$) and the performance of entanglement assisted classical communication ($\Delta_d^{Ent}$), achieved by maximal violation of the CGLMP inequality as discussed earlier. Furthermore, we also give the performance $\Delta_d^{ML}$ in $\mathbb{G}_d$ obtained from classical communication assisted by the stronger-than-quantum correlations obtained from imposing only the principle of Macroscopic Locality \cite{NW10} (the first NPA hierarchy level). In addition, a recent method to bound the quantum set of probability distributions by hierarchies of dimensionally constrained quantum correlations \cite{NV15} allows us to find also upper bounds on $\Delta_d$.  For the cases of $d=3,4$ which we have been able to study with the available computational resources, the upper bound coincides with the lower bound in table \ref{tab:1} up to a precision of order $10^{-11}$ and therefore strongly indictes optimality. When $d=5,6$ we only have lower bounds on $\Delta^{QS}_d$ but we still achieve the same as $\Delta_d^{Ent}$ with a high precision of order $10^{-10}$. Due to the results in Ref. \cite{TZ16}, it follows that there for any $d$ exists a transmission of a quantum system strategy that achieves $\Delta_d^{QS}=\Delta_d^{Ent}$. Our numerics strongly indicate that this equivalence is true for $d=2,\ldots , 6$.
		
However, our results unravel a puzzling feature: the equvalence between the performance in $\mathbb{G}_d$ with transmission of a quantum system or entanglement assisted classical communication is suddenly broken when $d\geq 7$. In fact the optimal measurements of Bob when $d=2,...,6$ in the transmission of a quantum system case are the same as those used to maximally violate the CGLMP inequality \cite{CGLMP02}:
		\begin{equation}\label{CGLMPmeas}
		|m_{j,b}\rangle=\frac{1}{\sqrt{d}}\sum_{l=0}^{d-1}\omega^{l(-j+\alpha_b)}|l\rangle
		\end{equation}
		with $\alpha_0=1/4$ and $\alpha_1=-1/4$, where $j\in\{0,...,d-1\}$ labels the outcome, $b\in\{0,1\}$ labels the measurement, and $\{|l\rangle\}$ is the computational basis. However, when the dimension is increased to $d=7$, something interesting happens; the performance achievable by transmission of a quantum system is higher than what can be obtained by entanglement assisted classical communication when playing $\mathbb{G}_d$. The optimal measurements of Bob are no longer in the basis \eqref{CGLMPmeas}, which are optimal for CGLMP inequality violation, and therefore are optimal for entanglement-assisted $\mathbb{G}_d$. See supplementary material for the optimal state preparations and measurements for $d=7$. 
		However, if we fix the measurements of Bob to projections onto states of the form \eqref{CGLMPmeas}, and run interior point optimization with SDPs, we will find no discrepancy between the performance of the two quantum resources. Dimension $d=7$ serves as a threshold after which our results indicate that transmission of a quantum system always outperforms entanglement assisted classical communication, at least up to $d=11$. Furthermore, our results show that for $d=9,10,11$, transmission of a quantum system is even outperforming classical communication assisted by the stronger-than-quantum correlations obtained from Macroscopic Locality.

		{\it Conclusions. --- } We studied the performance of entanglement assisted classical communication versus transmission of a quantum system in an family of communication complexity problems parameterized by the dimension of local Hilbert space. We were able to show that transmission of a quantum system can outperform, not only entanglement assisted classical communication but also classical communication assisted by correlations limited only by Macroscopic Locality. This result shows that transmission of a quantum system can be a very powerful resource for information processing, and that its relationship to entanglement assisted classical communication is far richer than what was previously known. Furthermore, we note that the performances found using transmission of a quantum system are not necessarily the optimal ones, but constitute lower bounds. This may mean that the effect could be even more pronounced.
		
		However, our most intriguing result is that $d=7$ serves as an unexpected dimensional threshold at which the quantitative equivalence between the two quantum resources disappears. A satisfying explanation of the origin of this phenomenon is still missing; What is so special in dimension 7? 
		
		Our work opens a new direction for research; to characterize the relationship between entanglement assisted classical communication and transmission of a quantum system. This would be interesting for foundational insights in the relation of spatial and sequential correlations in quantum theory. Furthermore, if much increased non-classicality  can be observed with one resource over the other, this is certainly of interest in quantum information theory since this could lead to further enhancements of quantum communication protocols.

		This work is supported by the Swedish research council, ADOPT, FNP programme TEAM, ERC grant QOLAPS, and NCN grant no. 2014/14/E/ST2/00020. MZ acknowledges award-grant COPERNICUS of FNP and DFG.

	\end{document}